\documentclass[showpacs,preprintnumbers,amsmath,amssymb,aps,twocolumn]{revtex4}

\usepackage{amsfonts}
\usepackage[dvips]{graphicx}
\usepackage{color}

\newcommand{\eqnref}[1]{Eqn.~\eqref{#1}}
\newcommand{\figref}[1]{Fig.~\ref{#1}}
\newcommand{\secref}[1]{Sec.~\ref{#1}}

\newcommand{\diffd}{\text{d}}
\newcommand{\e}[1]{\text{e}^{#1}}
\newcommand{\cmplxi}{\text{i}}
\newcommand{\bracetextsize}{\displaystyle}
\newcommand{\tr}{\operatorname{tr}}
\renewcommand{\vec}[1]{\mathbf{#1}}
\newcommand{\obrace}[2]{\overbrace{#1}^{\bracetextsize{#2}}}
\newcommand{\ubrace}[2]{\underbrace{#1}_{\bracetextsize{#2}}}
\newcommand{\perc}{\,\%}

\begin{document}
\preprint{APS/123-QED}
\title{A many-flavor electron gas approach to electron-hole drops}
\author{G.~J. Conduit}
\email{gjc29@cam.ac.uk}
\affiliation{Theory of Condensed Matter, Department of Physics,
University of Cambridge, Cavendish Laboratory, 19, J.~J. Thomson
Avenue, Cambridge, CB3~0HE, UK}
\date{\today}

\begin{abstract}
A many-flavor electron gas (MFEG) is analyzed, such as could be found in a
multi-valley semiconductor or semimetal. Using the re-derived polarizability for
the MFEG an exact expression for the total energy of a uniform MFEG in the
many-flavor approximation is found; the interacting energy per particle is shown
to be $-0.574447(E_{\text{h}}a_{0}^{3/4}m^{*3/4})n^{1/4}$ with $E_{\text{h}}$
being the Hartree energy, $a_{0}$ Bohr radius, and $m^{*}$ particle 
effective mass. The short characteristic length-scale of the MFEG motivates a
local density approximation, allowing a gradient expansion in the energy
density, and the expansion scheme is applied to electron-hole drops,
finding a new form for the density profile and its surface scaling properties.
\end{abstract}

\maketitle

\section{Introduction}\label{sec:Introduction}

\begin{figure}
 \includegraphics[width=\columnwidth]{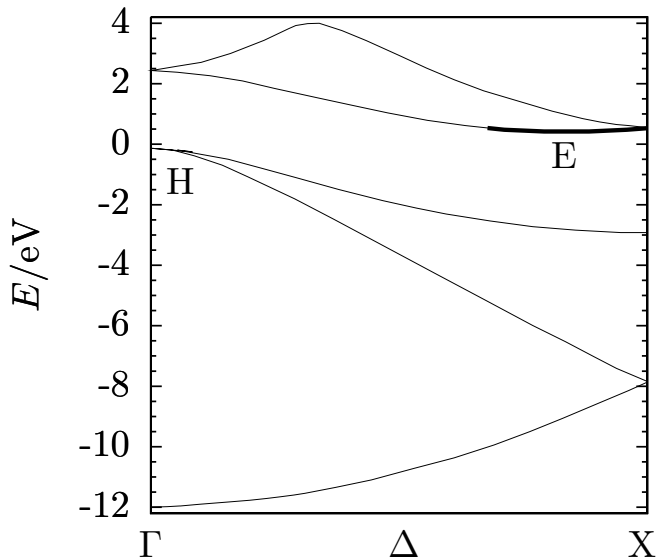}
 \caption{The Si band structure in the $[100]$ direction generated with a
 LDA-DFT approximation by a plane-wave pseudopotential method
 \cite{05csphprp05}. The Fermi energy is as $E=0~\text{eV}$; below are valence
 bands with holes at H; above conduction bands, the bold parabolic curve
 signifies the first conduction band valley with electrons at E.}
 \label{fig:SiBandStructure}
\end{figure}

For some semiconductors, at low temperatures and high density, electrons and
holes condense into electron-hole drops, which provide a good testing ground for
understanding effects of electron-electron interactions \cite{83jk10}. Some of
the semiconductors (and also semimetals) that electron-hole drops form in
\cite{76abkos08,76ko07}, such as Si, Ge, and diamond have conduction band minima
near the Brillouin zone boundary, for example Si has six degenerate valleys, see
\figref{fig:SiBandStructure}, a Ge-Si alloy has ten degenerate valleys, and
$\text{Pb}_{1-x-y}\text{Sn}_{x}\text{Mn}_{y}\text{Te}$ has twelve valleys in the
$\Sigma$ band \cite{90sksg12}. When the material is strained, valley degeneracy
reduces \cite{73br02,74vbs12,74ps07,74vds10,78kv03}, which can be experimentally
probed
\cite{78wmkfj08,78cw12,81wfs06,81gw11,82flwsr02,82cf11,89ssw08,00tstzszka04,05jwnk01,07nok11},
meaning that valley degeneracy could be regarded as a control parameter. Because
of this, as well as degeneracy being large in some semiconductors, valley
degeneracy might be a good parameter with which to formulate a theory of
electron-hole drops.

Previous theoretical analyses of electron-hole drops
\cite{73sss08,74r02,77r06,78mhl09,78m11,78kv03,79rly08} used an expansion of the
energy density with parameters found from separate energy calculations
\cite{74vds10}. An alternative approach is to assume that each valley contains a
different type of fermion, denoted by an additional quantum number, which we
shall call the \emph{flavor}, the total number of flavors (valleys) is $\nu$.
Further motivation to study flavors stems from the fact that in some previous
studies of multiply degenerate systems the number of flavors has not been well
defined, for example heavy fermions \cite{02zcja03,05k12,06bi02}, charged domain
walls \cite{98eogsz09}, a super-strong magnetic field \cite{76ko07}, and spin
instabilities \cite{86gq04,00mqg03}. Cold atom systems in optical lattices
\cite{04hh04,04hh09,07crd09} have a well defined number of flavors but weak
interactions between particles. In electron-hole drops however the number of
flavors is well defined and interactions are strong.

The ground state energy and pair correlation function of a free many-flavor
electron gas (MFEG) were examined using a numerical self-consistent approach for
the local field correction by \citet{94g08}, and superconductivity was studied
by \citet{64c04}. Following the method of \citet{76ko07}, \citet{76abkos08}
studied the behavior of the free MFEG by summing over all orders of Green's
function contributions, they found an exact expression for the correlation
energy of a MFEG (which dominates the interacting energy in the extreme
many-flavor limit). This paper describes the derivation of a more versatile
formalism, based on a path integral, which gives an exact expression for the
total energy of the MFEG; the theory could apply with as few as six flavors
where the exchange energy assumed small by \citet{76abkos08} would be
significant.

As well as studying the uniform case, the previously unstudied density response
of a MFEG not constrained to be uniform is investigated. The screening
length-scales of the MFEG are shown to be short relative to the inverse Fermi
momentum, suggesting that a local density approximation (LDA) might be a good
approximation, motivating a gradient approximation. This gradient expansion is
then applied to analyze the electron-hole drop density profile, and to simulate
effects of strain the scaling of drop surface thickness and tension with number
of flavors is examined.

In a MFEG with $\nu$ flavors at low temperatures the relationship between the
number density of electrons $n$ and Fermi momentum $p_{\text{F}}$ is
\begin{equation}
 \label{flavorsdensityofstates}
 n=\frac{\nu p_{\text{F}}^{3}}{3\pi^{2}}.
\end{equation}
When the electrons have multiple flavors, each Fermi surface encloses fewer
states so $p_{\text{F}}\propto\nu^{-1/3}$. The local band curvature governs the
electron effective mass, the band structure is often such that the holes relax
into a single valence band minimum at the $\Gamma$ point (see
\figref{fig:SiBandStructure}); here holes are assumed to be heavy and spread out
uniformly providing a jellium background.

The Thomas-Fermi approximation predicts a screening length $\kappa^{-1}=(4\pi
e^{2} g)^{-1/2}$, where $g$ is the density of states (DOS) at the Fermi
surface. The DOS is dependent on the number of flavors as
$g\propto\nu\sqrt{E_{\text{F}}} \propto \nu^{2/3}$ and so
$\kappa^{-1}\propto\nu^{-1/3}$. The ratio of the inverse
Fermi momentum length-scale to the screening length varies with number
of flavors as $p_{\text{F}}/\kappa\propto\nu^{-2/3}$. This paper takes the many
flavor limit $\nu\gg 1$, in which the screening length is smaller than the
inverse Fermi momentum, $\kappa^{-1}\ll p_{\text{F}}^{-1}$, the many-flavor
limit therefore means that the wave vectors of the strongest electron-electron
interactions obey $q\gg p_{\text{F}}$, this is the opposite limit to the RPA
which assumes that $p_{\text{F}}\gg\kappa$. Physically this means the
characteristic length-scales of the MFEG are short so a LDA can be used in
\secref{sec:AnalyticalGradient} to develop a gradient expansion.

\begin{figure}
 \includegraphics[width=\columnwidth]{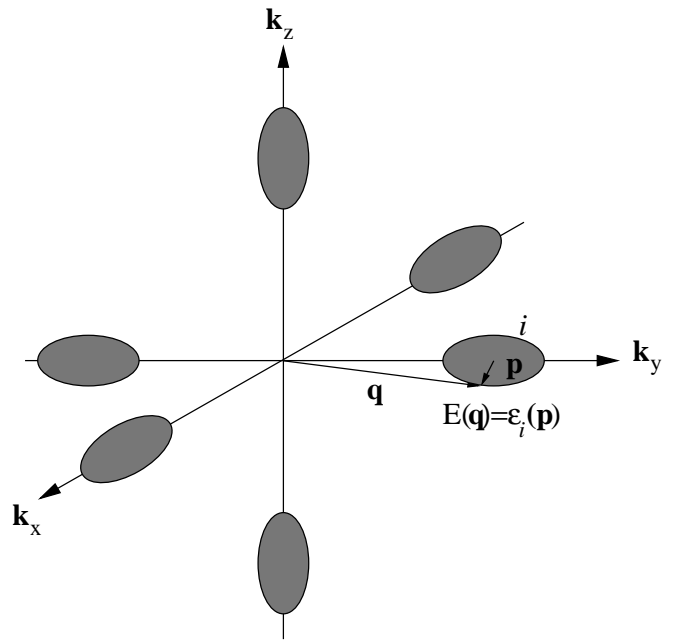}
 \caption{The dark grey ellipsoids show Fermi surfaces of electrons in the six
 degenerate conduction band valleys in Si. $E(\vec{q})$ is the energy with
 momentum $\vec{q}$ measured with respect to the $\Gamma$ point,
 $\epsilon_{i}(\vec{p})$ is energy with momentum $\vec{p}$ measured with respect
 to the center of the $i\text{th}$ valley.}
 \label{fig:SiMultiValley}
\end{figure}

The conduction band energy spectrum is characterized by two spectra $E(\vec{q})$
and $\epsilon_{i}(\vec{p})$ as shown in \figref{fig:SiMultiValley}. There are
two energy functions: $E(\vec{q})$ gives the energy in the band structure at
momentum $\vec{q}$; $\epsilon_{i}(\vec{p})=p^{2}/2m$ denotes the kinetic energy
at momentum $\vec{p}$ with respect to the center of the $i\text{th}$ valley (the
dispersions of all valleys are assumed to be the same and isotropic so that
$\epsilon_{i}(\vec{p})=\epsilon(\vec{p})$. \citet{76abkos08} have outlined a
method of calculating a scalar effective mass for anisotropic valleys).

Physical manifestations of the many-flavor limit include effects where the DOS
at the Fermi surface (energy $E_{\text{F}}$) is important; from
\eqnref{flavorsdensityofstates} the DOS of a particular flavor $i$ shrinks as
$g_{i}(E_{\text{F}})\propto p_{\text{F}}\propto\nu^{-1/3}$ whereas the DOS of
all flavors grows since $g(E_{\text{F}})=\sum_{i=1}^{\nu}g_{i}(E_{\text{F}})=\nu
g_{1}(E_{\text{F}})\propto\nu^{2/3}$. With increasing flavors, more electrons
are within $\sim k_{\text{B}}T$ of the Fermi surface hence are able to be
thermally excited, therefore the heat capacity of the MFEG,
$C=12k_{\text{B}}^{2}T(\nu/3\pi^{2}n)^{2/3}/5$, increases with number of
flavors. The Stoner criterion \cite{63k09,63h11} for band ferromagnetism states
that for opposite spin electrons interacting with positive exchange energy $U$,
ferromagnetism occurs when $g(E_{\text{F}})U\ge1$. With increasing number of
flavors the total DOS $g(E_{\text{F}})\propto\nu^{2/3}$ increases so that the
Stoner criterion becomes more favorable. However, this analysis does not take
into account the curvature of the DOS at the Fermi surface which can be an
important factor in determining whether ferromagnetism occurs
\cite{62wr11,93fld09}. The effect of the total DOS is also seen in the
paramagnetic susceptibility, this is proportional to the total DOS at the Fermi
surface so is expected to increase with number of flavors. Analogously one can
compare a transition metal with narrow $d$-bands that leads to a large DOS at
the Fermi surface with a simple metal that has broader free electron conduction
bands and so a lower DOS at the Fermi surface. Similar to many flavor systems,
transition metals are experimentally observed \cite{01am01} to have a
significantly higher specific heat capacity and greater magnetic susceptibility
than typical simple metals. The simple scaling relationships with number of
flavors for heat capacity and magnetization provide additional motivation to
analyze a MFEG in more detail.

This paper uses the atomic system of units, that is
$e^{2}=\hbar=m=1/(4\pi\epsilon_{0})=1$, but is modified so that $m$ denotes an
appropriate effective mass for the electron-hole bands, which is the same for
all valleys. This mass $m=m_{\text{e}}m^{*}$ can be expressed as a multiple of
the electron mass $m_{\text{e}}$ and the dimensionless effective mass
$m^{*}$. The units of length are then $a_{0}^{*}=a_{0}/m^{*}$ where $a_{0}$ is
the Bohr radius, units of energy are those of an exciton,
$E_{\text{h}}^{*}=E_{\text{h}}m^{*}$, where $E_{\text{h}}$ is the Hartree
energy. These six quantities, defined to be unity, give the standard atomic
units when $m^{*}=1$. For the important relationships that are derived in this
paper particular to the MFEG, the full units are shown explicitly for
clarity. Throughout this paper density is denoted by both $n$ (number density of
particles) and $r_{\text{s}}$ (Wigner-Seitz radius).

In this paper, firstly a new formalism for the uniform system is derived. In
\secref{sec:AnalyticalPolarizability} the system polarizability is found, in
\secref{sec:PartitionFunction} the general quantum partition function is
derived, and in \secref{sec:FreeCoulombGas} the uniform MFEG total energy is
calculated. Secondly, we examine the system with non-uniform density: in
\secref{sec:AnalyticalGradient} a gradient expansion in the density for the
total energy is found and is applied to electron-hole drops in
\secref{sec:ElectronHoleDrops}, whose density profile and surface properties are
calculated.

\subsection{Polarizability}\label{sec:AnalyticalPolarizability}

In this section the MFEG polarizability is derived. Though the result for the
polarizability is the same as previous work \cite{76abkos08,78br07}, the
derivation is presented here since an assumption made leads to an applicability
constraint on the many-flavor theory (in \secref{sec:DensityLimits}), and the
MFEG polarizability is an important quantity that will feature prominently in
two main results of this paper: the \secref{sec:AnalyticalResults} derivation of
the MFEG interacting energy and the derivation of the gradient expansion (see
\secref{sec:AnalyticalGradient}).

The polarizability $\Pi_{0}^{\text{MF}}(\vec{q},\omega)$, where the superscript
``MF'' (many-flavor) denotes this is only for a MFEG, at wave vector $\vec{q}$
and Matsubara frequency $\omega$ is given by the standard Lindhard form
\begin{equation}
 \Pi_{0}^{\text{MF}}(\vec{q},\omega)=\sum_{i,j=1}^{\nu}\frac{\delta_{i,j}}{4\pi^{3}}\int
 \frac{n_{{\text{F}}}\left(\epsilon_{i}\left(\vec{p}\right)\right)
 -n_{{\text{F}}}\left(\epsilon_{j}\left(\vec{p}+\vec{q}\right)\right)}
{\cmplxi\omega+\epsilon_{i}\left(\vec{p}\right)
-\epsilon_{j}\left(\vec{p}+\vec{q}\right)}\diffd\vec{p},
\end{equation}
where $n_{\text{F}}(\epsilon_{i})=1/(\e{\beta(\epsilon_{i}-\mu)}+1)$ is the
Fermi-Dirac distribution, $\beta=1/k_{\text{B}}T$, and $\mu$ is the chemical
potential. In the standard expression for the polarizability the Fermi-Dirac
distribution would contain the energy spectrum $E(\vec{p})$ but in the MFEG each
electron is in a particular valley so the polarizability should be re-expressed
in terms of the energy dispersion of each valley $\epsilon_{i}(\vec{p})$ (see
\figref{fig:SiMultiValley}) and the contributions must be summed over the
valleys $i$ and $j$. \eqnref{correlationenergysubstituted} shows that a large
Coulomb potential energy penalty $V(\vec{q})\propto1/q^{2}$ inhibits exchange
between different valleys so that the Kronecker delta $\delta_{i,j}$ removes
cross-flavor terms, and as all of the conduction valleys have the same
dispersion a factor of $\nu$ will replace the remaining summation over
valleys. Supposing each conduction valley has a locally quadratic isotropic
dispersion relationship (with effective mass $m$), symmetrizing results in
\begin{equation}
 \label{eqn:polarisabilitytwospheres}
 \Pi_{0}^{\text{MF}}=\frac{\nu}{4\pi^{3}}
 \int\frac{n_{\text{F}}\left(\frac{1}{2}\left|\frac{1}{2}\vec{q}
 -\vec{p}\right|^{2}\right)-n_{\text{F}}
 \left(\frac{1}{2}\left|\frac{1}{2}\vec{q}+\vec{p}\right|^{2}\right)}
 {\cmplxi\omega-\vec{p}\cdot\vec{q}} \diffd\vec{p}.
\end{equation}
In \secref{sec:FreeCoulombGas} it is shown that the typical momentum exchange
$q\sim(\hbar a_{0}^{-1/4})n^{1/4}$ is large relative to the Fermi momentum, therefore
the two volumes in momentum space of integration variable $\vec{p}$, defined by
the two Fermi-Dirac distributions are far apart relative to their radii
$q/2\gg p_{\text{F}}$, and the temperature is sufficiently low so that the
high energy tails of the two distributions have negligible overlap. Within these
approximations the simple form for the polarizability is
\begin{eqnarray}
 \label{refeqnpolarizabilityexpansion}
 \Pi_{0}^{\text{MF}}(\vec{q},\omega)=-\frac{n}{(\omega/q)^2+q^{2}/4},
\end{eqnarray}
this expression agrees with the many-flavor polarizability found by
\citet{76abkos08} and \citet{78br07}.

The standard Lindhard form for the polarizability, when taken in the same $q\gg
p_{\text{F}}$ limit as imposed by the many-flavor system agrees with
\eqnref{refeqnpolarizabilityexpansion}. In the static limit where
frequencies are small compared to the momentum transfer
$q^{2}\gg(\hbar^{3}/a_{0}^{*2}E_{\text{h}}^{*})\omega$, the polarizability
varies as $q^{-2}$. In this limit, one Green's function is restricted by the sum
over Matsubara frequencies to lie inside the Fermi surface whilst the other
gives the polarizability dependence of $1/\epsilon(\vec{q})\sim q^{-2}$ due to
the excited electron's kinetic energy.

The derivation of the polarizability accounted only for intra-valley scattering,
which means that in the MFEG the same terms contribute \cite{76abkos08} as in
the RPA for the standard electron gas. Therefore, diagrammatically, in the
polarizability all electron loops are empty, the polarizability contains only
reducible diagrams, which is denoted by the polarizability subscript ``0''.

\section{Analytic formulation}\label{sec:AnalyticalResults}

Having reviewed the derivation of the polarizability of the system it is now
used to formulate two complementary components of the many-flavor theory.  The
first is the derivation of the energy of a uniform system; we begin by
calculating the general quantum partition function in
\secref{sec:PartitionFunction} and continue for the homogeneous case in
\secref{sec:FreeCoulombGas}. The validity of the many-flavor approach for a
uniform MFEG is investigated in \secref{sec:DensityLimits}. The second part of the
formalism is a gradient expansion of the energy density, looked at in
\secref{sec:AnalyticalGradient}. Finally, the uniform and gradient expansion
parts of the formalism will be brought together to study the model system of
electron-hole drops in \secref{sec:ElectronHoleDrops}.

\subsection{Partition function}\label{sec:PartitionFunction}

To derive an expression for the total energy of the system a functional path
integral method is followed, which is a flexible approach that should be
extendable to investigate further possibilities such as modulated states and
inter-valley scattering. Fermion field variables $\psi$ are used to describe the
electrons irrespective of flavor in the dispersion $E(\vec{\hat{p}})$. Overall
the system is electrically neutral, so in momentum representation the
$\vec{q}=\vec{0}$ element is ignored. The repulsive charge-charge interaction
acting between electrons is $V(\vec{r})=e^2/r$, we explicitly include the
dependence on electron charge $e$ (even though it is defined to be unity) so
that the charge can be set equal to zero to recover the non-interacting
theory. For generality we consider stationary charges $Q(\vec{r})$ embedded in
the MFEG which have a corresponding static potential $U(\vec{r})$. The quantum
partition function for the MFEG written as a Feynman path integral is then
\begin{widetext}
 \begin{eqnarray}
  \mathcal{Z}&=&\iint\exp\left(\iint\bar{\psi}(\vec{r},\tau)
  \left(-\cmplxi\hat{\omega}+E\left(\vec{\hat{p}}\right)-\mu\right)
  \psi(\vec{r},\tau)\diffd\vec{r}\diffd\tau\right)\nonumber\\
  &\times&\exp\left(\frac{1}{2}
  \iiint\left(\bar{\psi}(\vec{r'},\tau)\psi(\vec{r'},\tau)
  -Q(\vec{r'})\right)V(\vec{r}-\vec{r'})
  \left(\bar{\psi}(\vec{r},\tau)\psi(\vec{r},\tau)-Q(\vec{r})\right)
  \diffd\vec{r}'\diffd\vec{r}\diffd\tau\right)
  \mathcal{D}\bar{\psi}\mathcal{D}\psi.
 \end{eqnarray}
 This expression for the quantum partition function differs from that used for
 an electron gas (which has just a single flavor) only by the operator
 $E(\vec{\hat{p}})$ which gives the appropriate energy dispersion. To recover the
 standard electron gas result, which has a free particle dispersion relationship
 centered at the $\Gamma$ point, one should set
 $E(\vec{p})=p^{2}/2m_{\text{e}}$. For the MFEG, as outlined in
 \figref{fig:SiMultiValley}, $E(\vec{p})$ represents the dispersion relationship
 of the whole conduction band, but no approximation concerning the flavors has
 yet been made, so the formalism applies for any number of flavors with a
 suitable energy dispersion relationship.

 To make the action quadratic in the fermion variable $\psi$, the
 Hubbard-Stratonovich transformation \cite{78k11} introduces an auxiliary boson
 field $\phi(\vec{r},\tau)$
 \begin{eqnarray}
  \label{refeqnCompleteQuantumPartitionFunction}
  \mathcal{Z}&=&\int\exp\Biggl(-\frac{\beta\Omega}{2}
  \sum_{\vec{q}\ne\vec{0},\omega} \phi(\vec{q},\omega) (q^2/4\pi)
  \phi(-\vec{q},-\omega)+ \obrace{\frac{\pi e^{2}\beta
  n}{2}\sum_{\vec{q}\ne\vec{0}}\frac{4\pi}{q^{2}}}{\dagger}\Biggr)
  \nonumber\\ &\times&\iint\exp\left(\iint\bar{\psi}(\vec{r},\tau)
  \left(-\cmplxi\hat{\omega}+E\left(\vec{\hat{p}}\right)+\hat{U}-\mu
  +\cmplxi e\hat{\phi}\right)\psi(\vec{r},\tau)
  \diffd\vec{r}\diffd\tau\right)\mathcal{D}\bar{\psi}\mathcal{D}
  \psi\mathcal{D}\phi.
 \end{eqnarray}
\end{widetext}
The direct decoupling channel \cite{78k11} was chosen as the relevant
contributions come from a RPA-type contraction of operators.

The term labeled with a $(\dagger)$ exclusive of both fermion
variables $\psi$ and the auxiliary field $\phi$, physically removes the
electron self-interaction included
when expressing the auxiliary field in a Fourier
representation. Integrating over the fermion variables $\psi$ and
using ${\ln}({\text{det}}\hat{A})={\tr}({\ln}\hat{A})$ gives
$\mathcal{Z} = \int \e{-S[\phi]} \mathcal{D}\phi$, where the action
$S[\phi]$ is
\begin{eqnarray}
 \label{refeqnGeneralAction}
 S[\phi]&=&\frac{\beta\Omega}{2\pi} \sum_{\vec{q}\ne\vec{0},\omega}
 \phi(\vec{q},\omega)(q^2/4\pi) \phi(-\vec{q},-\omega)\nonumber\\
 &-&2\pi e^{2}\beta
 n\sum_{\vec{q}\ne\vec{0}}\frac{1}{q^{2}}\nonumber\\
 &-&{\tr}\bigl({\ln}\bigl(\ubrace{-\cmplxi\hat{\omega}
 +E\left({\vec{\hat{p}}}\right)+\cmplxi
 e\hat{\phi}+\hat{U}-\mu}{\hat{G}_{\phi}^{-1}}\bigr)\bigr).
\end{eqnarray}
Due to its similarity to an inverse Green function,
$\hat{G}_{\phi}^{-1}$ is used to denote the argument of the
logarithm, the subscripts ``$\phi$'' or ``$0$'' denote
whether the inverse Green's function includes the auxiliary
field or is free.

Finally, we note for use later that the ground state total energy per particle
$E_{\text{G}}=E_{\text{int}}+E_{0}$ can be split into two components. The
interacting energy is $E_{\text{int}}$ (found in \secref{sec:FreeCoulombGas})
and the non-interacting energy is
\begin{equation}
 \label{refeqnKineticEnergy}
 E_{0}=\frac{3}{10}\left(3\pi^{2}\frac{n}{\nu}\right)^{2/3},
\end{equation}
which is the energy with interaction between charges switched off $(e=0)$. It
falls with increasing number of electron flavors due to the shrinking Fermi
surface.

\subsection{Homogeneous Coulomb gas}\label{sec:FreeCoulombGas}

So far, up to \eqnref{refeqnGeneralAction}, the formalism is exact, however to
perform the functional integral over bosonic variable $\phi$ an approximation
must be made. To proceed one notes that with no external potential
$U(\vec{r})=0$ the saddle-point auxiliary field of the action
(\eqnref{refeqnGeneralAction}) is $\phi=0$, fluctuations in the action are
expanded about the saddle point solution in $\phi$ giving the expression
\begin{eqnarray}
 \label{eqn:ActionExpansionInPhi}
 S[\phi]&=&\ln(\hat{G}_{0}^{-1})
 +\tr\Bigl(\hat{\phi}\hat{V}^{-1}\hat{\phi}-\frac{1}{2}\obrace{\hat{\phi}\hat{G}_{0}
 \hat{\phi}\hat{G}_{0}}{\ddagger}\Bigr) \nonumber\\
 &-&\frac{1}{4}\tr\left(\hat{\phi}\hat{G}_{0}\hat{\phi}\hat{G}_{0}\hat{\phi}
 \hat{G}_{0}\hat{\phi}\hat{G}_{0}\right)
 +\mathcal{O}(\phi^{6})\nonumber\\&-&\frac{e^{2}}{2}\beta
 n\sum_{\vec{q}\ne\vec{0}}\frac{4\pi}{q^{2}}.
\end{eqnarray}
Terms are now kept to quadratic order in $\phi$, this is equivalent to the RPA,
analogous to the terms kept in the derivation of the many-flavor polarizability,
see \secref{sec:AnalyticalPolarizability}. This approximation will place a
constraint on the validity of the formalism that is further examined in
\secref{sec:DensityLimits}.

The product of two Green's functions in the quadratic term in $\phi$, labeled
$(\ddagger)$, is identified with the polarizability $\Pi_{0}$, this is still
expressed in terms of the general energy spectrum $E(\vec{p})$ so is not yet
necessarily many-flavor and does not carry the superscript ``MF'' used in
\secref{sec:AnalyticalPolarizability}. Following a multi-dimensional Gaussian
integral over the fluctuating field $\phi$ (to quadratic order) the quantum
partition function is
\begin{equation}
 \label{refeqnFreeCoulombGasPartitionFunction}
 \mathcal{Z}= \prod_{\vec{q},\omega}
 \left(q^{2}/4\pi-e^{2}\Pi_{0}(\vec{q},\omega)\right)^{-\frac{1}{2}}
 \exp\left(\frac{\beta}{2}\sum_{\vec{q}\ne\vec{0}}\frac{4\pi e^{2}}{q^{2}}n\right).
\end{equation}
In the low temperature limit we consider the free energy to get
$E_{\text{G}}=-{\text{lim}}_{\beta\rightarrow\infty}
({\ln}(\mathcal{Z})/\beta)=E_{\text{0}}+E_{\text{int}}$ to get the interacting
energy per particle, normalized so that $E_{\text{int}}=0$ with no interactions
$(e=0)$,
\begin{eqnarray}
 \label{eqn:GeneralInteractingEnergy}
 E_{\text{int}}&=&\frac{1}{2n}\Biggl(\iint{\ln}\left(1-\frac{4\pi
 e^{2}}{q^{2}}\Pi_{0}(\vec{q},\omega)\right)
 \frac{\diffd\omega\diffd\vec{q}}{(2\pi)^{4}}\nonumber\\
 &-&e^{2}n\sum_{\vec{q}\ne\vec{0}}\frac{4\pi}{q^{2}}\Biggr).
\end{eqnarray}
This equation remains general and is not necessarily in the many-flavor limit,
it is in agreement with previous expressions for the interacting energy
\cite{73br02} studied not in the many-flavor limit, but which use alternative
forms for the polarizability. If the standard (single flavor) electron gas form
for the polarizability, the Lindhard function, is used then it is possible to
recover, in the high density limit, the Gell-Mann Br\"uckner \cite{57gb04}
expression for the total energy.

However, to proceed, one should now assume many-flavors and use the appropriate
polarizability, \eqnref{refeqnpolarizabilityexpansion}. The many-flavor
polarizability summed over all Matsubara frequencies in the zero temperature
limit $\beta\rightarrow\infty$ satisfies
$\frac{1}{\beta}\sum_{\omega}\Pi_{0}^{\text{MF}}(\vec{q},\omega)=-n$. This is
used to substitute for the electron density $n$ in the final term in
\eqnref{eqn:GeneralInteractingEnergy} to yield the many-flavor result
\begin{eqnarray}
 \label{correlationenergy}
 E_{\text{int}}&=&\frac{1}{2n}\iint\left(\ln\left(1-\frac{4\pi
 e^{2}}{q^{2}}\Pi_{0}^{\text{MF}}(\vec{q},\omega)\right)\right.\nonumber\\
 &+&\left.\frac{4\pi e^{2}}{q^2}\Pi_{0}^{\text{MF}}(\vec{q},\omega)\right)
 \frac{\diffd\omega\diffd\vec{q}}{(2\pi)^{4}}.
\end{eqnarray}

To evaluate the interacting energy one first substitutes for the many-flavor
polarization using \eqnref{refeqnpolarizabilityexpansion}, makes the change of
variables $\Omega=\omega/q^{2}$ and $Q=q/n^{1/4}$, and re-arranges to get
\begin{eqnarray}
 \label{correlationenergysubstituted}
 &&E_{\text{int}}=-n^{1/4}\nonumber\\
&&\times\ubrace{\frac{1}{(2\pi)^{3}}\iint\frac{16\pi e^{2}}{1+4\Omega^{2}}-Q^{4}\ln\left(1+\frac{16\pi e^{2}/Q^{4}}{1+4\Omega^{2}}\right)\diffd\Omega\diffd Q}{A_{\text{3D}}}.\nonumber\\
\end{eqnarray}
The integral is independent of density and number of flavors, so is the
numerical factor
$A_{3\text{D}}=(E_{\text{h}}^{*}a_{0}^{*3/4})\Gamma(-5/4)\Gamma(3/4)/(2\pi^{5/4})\approx0.574447(E_{\text{h}}^{*}a_{0}^{*3/4})$
that was evaluated analytically \footnote{This differs from the value reported
by \citet{76abkos08} and \citet{76ko07} of
$A_{3\text{D}}=32(2\pi)^{3/4}2^{1/2}(E_{\text{h}}^{*}a_{0}^{*3/4})/(5\Gamma^{2}(1/4))$
by a factor of $2^{9/4}$. The result presented here was confirmed by three
separate methods: analytically, numerically, and by comparing with the initial
results of QMC simulations on the many-flavor system \cite{08gjc}.}. The
interacting energy is therefore
\begin{equation}
 \label{refeqncorrelationenergy}
 E_{\text{int}}=-A_{\text{3D}}n^{1/4},
\end{equation}
which is independent of the number of flavors. In evaluating
\eqnref{correlationenergysubstituted} the main contribution to the integral over
$Q=q/n^{1/4}$ is at a momentum $q\propto(\hbar a_{0}^{*-1/4})n^{1/4}$, so the
interaction and screening length-scale in a MFEG is $\lambda\sim\hbar/q\propto
a_{0}^{*1/4}n^{-1/4}\ll\hbar/p_{\text{F}}$, which is shorter than the Fermi
momentum length-scale.

The interacting energy $E_{\text{int}}=E_{\text{ex}}+E_{\text{corr}}$ can be
split into exchange energy $E_{\text{ex}}$ and correlation energy
$E_{\text{corr}}$. The interacting energy is independent of number of flavors,
the exchange energy, $E_{\text{ex}}=-(3/2)(3n/\pi\nu)^{1/3}$ \cite{94g08}, falls
with number of flavors, therefore the correlation energy dominates over the
exchange energy in the interacting energy in the many-flavor limit. In terms of
the total energy, the correlation energy also dominates over the non-interacting
energy, the kinetic energy that falls with number of flavors as
$E_{0}\propto\nu^{-2/3}$.  The increasing importance of the correlation energy
can be understood further by considering the electron pair correlation
function. With increasing number of flavors the length-scales between electrons
of the same flavor increase as $\propto\nu^{1/3}r_{\text{s}}$ and thus exchange
energy and kinetic energy reduce whereas the correlation energy depends only on
the distance $r_{\text{s}}$ between electrons so is unaffected by the number of
flavors present.  \citet{76abkos08} and \citet{76ko07} found the
\eqnref{refeqncorrelationenergy} expression to be the correlation rather than
interacting energy, neglecting the exchange energy which is small in the extreme
many flavor limit. In \secref{sec:DensityLimits} the expression for the
interacting energy is compared with self-consistent numerical calculations
\cite{94g08} on a MFEG with up to six flavors.

The interacting energy of a standard electron gas with a single flavor 
\cite{94g08} is more negative than that of a MFEG, which in turn is more 
negative than that of a Bose condensate \cite{92g02}, this could be due 
to the reducing negativity of the exchange energy, important in the single 
flavor system, but zero in the Bose condensate. In these two extreme systems, 
the single flavor electron gas and the Bose condensate, there is no notion 
of valley degeneracy, and therefore the intermediate system, the MFEG,
might be expected at most to have only a weak dependence on number of valleys. 
In fact, the interacting energy of the MFEG, over the range of density 
found in \secref{sec:DensityLimits}, contains no 
dependence on the number of valleys. The absence of flavor dependence is 
also present in the universal behavior for the exchange-correlation energy 
in electron-hole liquids proposed by \citet{82vk05}. 

The non-interacting energy term $E_{0}\propto (n/\nu)^{2/3}$ favors low electron
density, the interacting term $E_{\text{int}}= -A_{\text{3D}}n^{1/4}$ favors
high electron density, therefore the total energy per particle has a minimum as
a function of density of $E_{\text{Gmin}}\propto-\nu^{2/5}$ at
$n_{\text{min}}\propto\nu^{8/5}$. The presence of a minimum in energy with
density of the MFEG is consistent with the results of \citet{76abkos08} and
\citet{73br02} who analyzed conduction electrons in a semiconductor. One
consequence of this minimum is the possibility of a low density phase coexisting
with excitons.

Before analyzing the non-uniform system in detail in
\secref{sec:AnalyticalGradient} we can make qualitative arguments about its
expected behavior within a potential well. According to Thomas-Fermi theory, an
electron gas in a slowly varying attractive potential has a constant chemical
potential. The electron gas is least dense at the edges of the potential and is
densest at the center of the well. In a MFEG, due to the negative interacting
energy $E_{\text{int}}= -A_{\text{3D}}n^{1/4}$ favoring high electron density,
the density is expected to further reduce at the edges of the attractive
potential and increase at the center of the well. In a repulsive potential the
opposite should occur.

\subsection{Density limits}\label{sec:DensityLimits}

In this section we will derive approximate expressions for the upper and lower
density limits over which the many-flavor limit applies, these will be used to
check the theory against numerical results \cite{94g08} and to predict a lower
bound on the number of flavors required for the theory to apply.

To find the upper density limit one notes that
\eqnref{correlationenergysubstituted} implies that an acceptable upper limit to
the momentum integral would scale as $q=\alpha(\hbar a_{0}^{*-1/4})n^{1/4}$, the
constant $\alpha\approx4$ was determined numerically and was chosen to give the
$q$ upper limit on the integral that recovered $95\perc$ of the interacting
energy. Additionally, the two regions of integration defined by the Fermi Dirac
distributions in \eqnref{eqn:polarisabilitytwospheres} must not overlap,
requiring that $q/2\ge p_{\text{F}}$. Combining these requires that for the
many-flavor limit to apply the density must satisfy
$na_{0}^{*3}\ll(\alpha^{12}\nu^{4})/(2^{12}3^{4}\pi^{8})$. Physically the
breakdown at high density is due to the strongest interactions taking place on
length-scales longer than the inverse length $p_{\text{F}}^{-1}$.

The low density limit is derived by considering the expansion of the action in
the auxiliary boson field $\phi$, \eqnref{eqn:ActionExpansionInPhi}. In order to
evaluate the Gaussian functional integral over the bosonic variable $\phi$ it is
necessary to neglect the quartic term in $\phi$, valid only when investigating
the system with respect to its long-range behavior, that is
$p_{\text{F}}\hbar a_{0}^{*}\gg1$ \cite{78k11}, and therefore
$na_{0}^{*3}\gg\nu/3\pi^{2}$. The breakdown at low density can be understood
because the MFEG is effectively a boson gas, all electrons will be in the
$\Gamma$ state ($\vec{k}=\vec{0}$) and there is no exchange energy.

The upper and lower critical density limits can be combined to conclude that the
many-flavor limit result for interacting energy applies for densities that obey
$0.03\nu\ll na_{0}^{*3}\ll0.005\nu^{4}$; this density range increases as
$\nu^{4}$ with number of flavors, the scaling relationship is the same as the
applicable density range of the correlation energy found by \citet{76abkos08},
though they did not provide estimates of numerical factors.

Using the above high and low density limits it is possible to estimate the
minimum number of flavors required for the theory to apply.  This is done by
setting the lower and upper estimates for the allowable density to be equal,
which gives $\nu\gtrapprox2$, this estimate is approximate due to the possible
inaccuracies in the upper and lower critical densities used in its derivation.
As the upper and lower critical densities have been set equal, the many-flavor
theory will apply here only over a very narrow range of densities, but this
range widens with increasing number of flavors as $\nu^{4}$. An alternative
limit can be found by comparing the interaction energy predicted by the theory
over the expected density range of applicability with the results of
\citet{94g08}. Their numerical self-consistent approach gives interaction
energies accurate to approximately $3\perc$ when compared with single-flavor
electron gas quantum Monte Carlo (QMC) calculations \cite{78c10,80ca08} and some
initial many-flavor QMC calculations \cite{08gjc}. At two flavors the
interacting energy predicted by the many-flavor theory is $\sim10\perc$ more
positive than the self-consistent numerical results \cite{94g08} indicating the
many-flavor theory does not apply at two flavors. For six flavors over the
predicted allowed density range the many-flavor theory is between $\sim0\perc$
and $\sim4\perc$ more positive than the numerical results, which indicates that
the many-flavor theory can be applied within the predicted range of
applicability ($0.5<r_{\text{s}}/a_{0}^{*}<1$). The theory should be applicable
in common multi-valley compounds, such as silicon which has six conduction band
valleys, and to those with more valleys \cite{90sksg12}. This result is
corroborated by the results of initial QMC calculations \cite{08gjc} on systems
with between 6 and 24 flavors.

In the first half of this paper a new versatile formalism to describe
a MFEG has been developed that could apply in systems containing approximately 
six or more degenerate conduction valleys. An exact expression for the 
total energy of the uniform MFEG was found and the applicable
density range derived. The next step is to investigate the response of
the MFEG to an external potential. A gradient approximation is
developed in \secref{sec:AnalyticalGradient} and this is applied to
electron-hole drops in \secref{sec:ElectronHoleDrops}.

\section{Gradient correction}\label{sec:AnalyticalGradient}

In \secref{sec:Introduction} it was shown that the typical length-scales of the
MFEG are short $q\gg p_{\text{F}}$, motivating a local density approximation
(LDA). This motivation is in addition to the usual reasons for the success of
the LDA in density-functional theory (DFT) \cite{92ptaaj10} -- that the LDA
exchange-correlation hole need only provide a good approximation for the
spherical average of the exchange-correlation hole and obey the sum rule
\cite{89jg06}. In this section the LDA is used with the polarizability derived
in \secref{sec:AnalyticalPolarizability} to develop a gradient correction to the
energy density that allows the theory to be applied to a non-uniform MFEG.

The typical momentum transfer in the MFEG is $q\sim(\hbar a_{0}^{*-1/4})n^{1/4}$
hence the shortest length-scale over which the LDA may be made is approximately
$(a_{0}^{*1/4}/\hbar)n^{-1/4}$ and the maximum permissible gradients in electron
density are $|\nabla n|_{\text{max}}\sim qn\sim(\hbar
a_{0}^{*-1/4}))n^{5/4}$. The gradient expansion will break down for short scale
phenomena, for example a Mott insulator transition. To derive an energy density
gradient expansion we follow \citet{64hk11} and \citet{74r02} and consider an
external charge distribution $n_{\text{ext}}(\vec{q})$ that couples to the
induced charge distribution $n_{\text{ind}}(\vec{q})$ with Coulomb energy
density
\begin{equation}
 -\frac{1}{2}\sum_{\vec{q}}\frac{4\pi
 e^{2}}{q^{2}}n_{\text{ext}}(\vec{q})n_{\text{ind}}(\vec{q}).
\end{equation}
One now substitutes for $n_{\text{ext}}(\vec{q})$ using the relative
permittivity
$1/\epsilon(\vec{q})=1+n_{\text{ind}}(\vec{q})/n_{\text{ext}}(\vec{q})=1/(1-4\pi\Pi_{0}^{\text{MF}}/q^{2})$
and the many-flavor polarizability \eqnref{refeqnpolarizabilityexpansion}. The
highest order term in $1/q^{2}$ gives the induced charge Coulomb energy, the
term of order $q^{2}$ is associated with a gradient expansion, in real space
this gives the expansion for the total energy per particle
\begin{equation}
 \label{eqn:ManyFlavorGradientExpansion}
 E_{\text{G}}+\frac{(\nabla n)^{2}}{8n^{2}}+\mathcal{O}\left((\nabla
 n)^{4}\right),
\end{equation}
here $E_{\text{G}}$ is ground state energy of the uniform system found in
\secref{sec:FreeCoulombGas}. The form of the energy correction is similar to the
von Weizs\"acker term \cite{35w07}, although here it is larger, having a
coefficient of $1/8$ rather than $1/72$ as in the von Weizs\"acker case. The
difference can be qualitatively understood by considering the Fermi surfaces
involved in the two cases for a given wave vector $q$, in the many-flavor case
the Fermi surfaces involved in the integral of
\eqnref{eqn:polarisabilitytwospheres} do not overlap as $q/2\gg p_{\text{F}}$ so
there is a large volume in Fermi space available hence a large coefficient of
$1/8$, whereas in the ordinary electron gas (single flavor) the same Fermi
surfaces do overlap, as now $q/2\ll p_{\text{F}}$, reducing the volume available
for integration so reducing the coefficient to $1/72$.

The gradient correction for the energy could be used in analytical
approximations or as a DFT functional. This energy density expansion allows the
MFEG to be applied to a variety of systems, its use for studying electron-hole
drops is demonstrated in \secref{sec:ElectronHoleDrops}.

\section{Electron-hole drops}\label{sec:ElectronHoleDrops}

In this section the MFEG is applied to a simple system to investigate the
properties of, electron-hole drops.  An electron-hole drop is a two-phase system:
a spherical region of a MFEG surrounded by an exciton gas \cite{78br07}. The
density profile, surface thickness and surface tension of drops are
investigated; the scaling of surface thickness and tension with number of
flavors is also found since this is can be experimentally probed through
externally imposed strain reducing the valley degeneracy \cite{74ps07}.

There have been four main theoretical methods used to analyze an electron-hole
drop in silicon and germanium, semiconductors which have six and four flavors
respectively. \citet{77r06} fitted an analytic form to the energy density
minimum and included the lowest order of a local gradient correction, from the
equation for energy density an analytic form for the density profile was
derived. A similar approach was used by \citet{73sss08} and \citet{74r02} to
study the surface structure in more detail. The second approach
\cite{78mhl09,78m11}, which was also applicable to situations with an external
magnetic field and uniaxial strain, conserved momentum, particle number and
pressure balance at the drop surface, the resulting equations were then solved
numerically. A third approach followed by \citet{78kv03} used a Pad\'e
approximant for the energy density \cite{74vds10} derived specifically for
silicon and germanium but did not include a gradient correction factor. The
fourth approach of \citet{79rly08} again used a Pad\'e approximant for the
energy density and also included a gradient correction factor. The latter two
approaches assumed an exponential density profile for the drop. These four
methods all use approximate forms for the energy density, an advantage of the
many-flavor approach is that the exact form for the analytic energy density
(within the many flavor assumption) can be used to solve for the drop density
profile.  Whilst analytic forms for the inner and outer density profile as well
as a model for the entire profile can be derived, the general problem must be
solved numerically. With an exact form for the density profile, electron-hole
drop surface effects can be studied.

Local charge neutrality is assumed so that the density of electrons and holes
are everywhere identically equal. A LDA with gradient correction is used so the
drop energy density is written as the sum of the local non-interacting, local
interacting and the lowest order term in a gradient expansion,
\begin{equation}
 \varepsilon(\vec{r})=\frac{3}{10}
 \left(\frac{3\pi^{2}}{\nu}\right)^{2/3}n(\vec{r})^{5/3}
 -A_{3\text{D}}n(\vec{r})^{5/4}+\frac{(\nabla
 n(\vec{r}))^{2}}{8n(\vec{r})}.
\end{equation}
The total energy of a drop is $\int\varepsilon(\vec{r})\diffd\vec{r}$ and the
total number of electrons in the drop is $\int n(\vec{r})\diffd \vec{r}$. The
total energy is minimized with respect to electron density $n(\vec{r})$ whilst
keeping a constant number of electrons in the drop by applying the
Euler-Lagrange equation with a Lagrange multiplier $\mu$, which represents the
chemical potential. If the drop has spherical symmetry the density must satisfy
\begin{eqnarray}
 \label{eqn:dropdiffeqn}
 &&2rn\frac{\diffd ^{2}n}{\diffd r^{2}}+4n\frac{\diffd n}{\diffd
 r}-r\left(\frac{\diffd n}{\diffd r}\right)^{2}\nonumber\\
 &&=16\left(\frac{3\pi^{2}}{\nu}\right)^{2/3}rn^{8/3}
 -40A_{3\text{D}}rn^{9/4}-32\mu rn^{2}.
\end{eqnarray}
The boundary conditions are specified at the center of the drop, where the
density takes the equilibrium homogeneous MFEG value and the density is smooth,
namely $n(0)=\bar{n}$ and $n'(0)=0$. The differential equation
\eqnref{eqn:dropdiffeqn} cannot be solved analytically for $n(\vec{r})$, but a
solution, $n(\vec{r})=\bar{n}$, exists for $\mu=0$ which corresponds to the
homogeneous MFEG, that is a drop containing an infinite number of
electrons. Before solving the differential equation numerically, two approximate
schemes are developed, one that applies near the drop center and the other near
the drop edge, and their predictions are compared with existing density profile
forms.

Near the center of the drop a perturbation solution about the equilibrium
density, $n(\vec{r})=\bar{n}+\Delta n(\vec{r})$ where $\Delta
n(\vec{r})\ll\bar{n}$, is considered. The solution to \eqnref{eqn:dropdiffeqn}
for the density is then
\begin{equation}
 \label{eqn:ehDropDensityInner}
 n(r)=\bar{n}+\frac{8\mu\bar{n}}{Q^{2}}\left(1-\frac{\text{sinh}(Qr)}{Qr}\right).
\end{equation}
This density profile is characterized by an exponential reduction of the density
away from $\bar{n}$ at the center.  The energy $Q^{2}$ is physically the rate of
change of energy per unit volume with respect to changing particle density, with
\begin{equation}
 \label{eqn:ehDropQ2Defn}
 Q^{2}=\frac{64}{3}\left(\frac{3\pi^{2}}{\nu}\right)^{2/3}
 \bar{n}^{2/3} -45A_{3\text{D}}\bar{n}^{1/4}-32\mu.
\end{equation}

The second approximation scheme applies in the drop tail where electron density
is low, $n(\vec{r})\ll\bar{n}$. The term containing the chemical potential is
disregarded as it is arbitrarily small for the large drops under investigation,
the non-interacting and interacting energy terms contain higher powers of
density so are negligibly small. In this regime the solution to
\eqnref{eqn:dropdiffeqn} is
\begin{equation}
 n(r)=
 \begin{cases}
  n_{0}\left(\frac{1}{r} -\frac{1}{r_{0}}\right)^{2} &r<r_{0},\\
  0&r\ge
  r_{0}.
 \end{cases}
 \label{eqn:ehDropDensityOuter}
\end{equation}
Here $n_{0}$ and $r_{0}$ are variational parameters which must be fitted to a
numerical solution.  This analytic form shows that the electron-hole drop has a
definite outer radius $r_{0}$, which is approached parabolically, it is also
noted that in the drop tail the solution obeys the differential equation
\begin{equation}
 \frac{1}{r^{2}}\frac{\diffd}{\diffd r}\left(r^{2}\frac{\diffd
 n^{1/2}}{\diffd r}\right)=\nabla^{2}n^{1/2}=0.
\end{equation}
If electron density is mapped onto a wave function $\psi$ through $n=|\psi|^{2}$
then the solution \eqnref{eqn:ehDropDensityOuter} obeys Schr\"odinger's equation
at low energy, that is $\nabla^{2}\psi=0$. The implied Schr\"odinger equation is
for a low density MFEG with negligible interaction between electrons due to
their large separation, consistent with the original assumption of low density
in the drop tail.

Previous studies of electron-hole drops \cite{77r06,78kv03,79rly08} had a
solution with the same exponential form both inside and outside of the drop, our
inner functional form, an exponential, agrees with previous work
\cite{77r06,78kv03,79rly08}, but our outer functional form, a quadratic-like
polynomial, does not agree with the exponential decay seen in previous work.
However, at the outside of the drop density is low and the arguments of
\secref{sec:DensityLimits} show the many flavor theory, which requires that the
density satisfies $n\gg0.03\nu$, does not apply here. The other theories
\cite{77r06,78kv03,79rly08} also do not apply in the low density region so both
the many-flavor and previous theories fail to agree only where they are not
applicable.

\begin{figure}
 \includegraphics[width=\columnwidth]{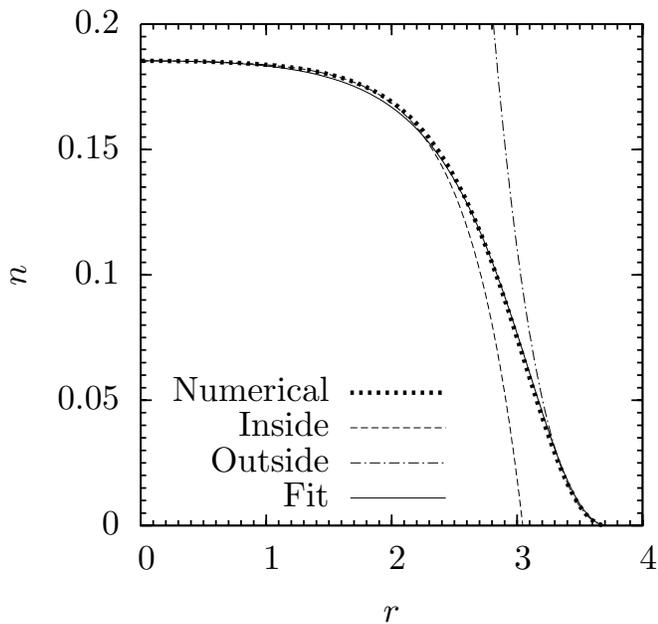}
 \caption{The density profile of a 12 flavor electron-hole drop with density
 parameter $r_s=1$. The numerical solution is shown by the dotted line,
 analytical approximations to the inside (outside) of the drop by the dashed 
 (dot-dashed) lines, and a best fit model fitted to the numerical solution by the solid
 line.}
 \label{ehDensityProfile}
\end{figure}

Using just the solution for the density in the drop tail
\eqnref{eqn:ehDropDensityOuter}, a reasonable analytical approximation for the
density form of the whole drop is
\begin{equation}
 n(r)=\left(\frac{1}{n_{0}(1/r-1/r_{0})}+\frac{1}{\bar{n}}\right)^{-1}.
 \label{eqn:ehDropDensityModel}
\end{equation}
This solution has the correct functional form at both the inside
($n(r)\rightarrow\bar{n}$) and outside of the drop and extrapolates smoothly in
between. It can be fitted to the actual solution using parameters $n_{0}$ and
$r_{0}$. However, the general differential equation is solved numerically giving
the density profile shown in \figref{ehDensityProfile}. The numerical solution
is well approximated in the inner and outer regions by
\eqnref{eqn:ehDropDensityInner} and \eqnref{eqn:ehDropDensityOuter}
respectively, and the model \eqnref{eqn:ehDropDensityModel} provides a good fit to
the numerical solution, having just a slightly too shallow gradient around the
median density but it agrees at both the center and outside of the well.

To allow us to compare properties of electron-hole drops predicted using
many-flavor theory with other work \cite{78kv03,79rly08}, one can characterize
the electron-hole drop properties through its surface thickness $D$ and tension
$\gamma$. The surface thickness $D$ is the width over which the density falls
from $90\perc$ to $10\perc$ of its homogeneous equilibrium value $\bar{n}$. The total
surface energy is the difference between the energy per unit area of the MFEG in
the drop and the energy of the same number of particles at equilibrium density
in a homogeneous system. The surface tension $\gamma$ is the total surface
energy divided by the characteristic drop surface area, here taken to be the
area of the spherical surface at the median density, which corresponds to a
characteristic drop radius $r_{\text{m}}$.
\begin{figure}
 \includegraphics[width=\columnwidth]{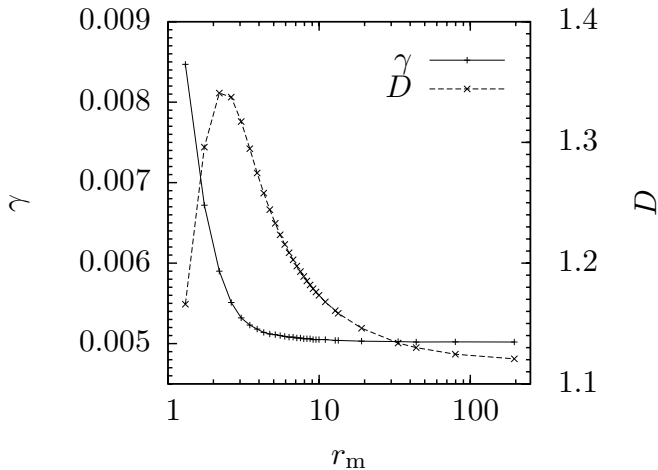}
 \caption{The surface tension $\gamma$ of the 12 flavor drop of radius
 $r_{\text{m}}$ is shown using the solid line and pluses based on the left-hand
 y-axis. The variation of the surface thickness $D$ is shown using the dashed
 line and crosses based on the right-hand axis, each point represents a separate
 simulation.}
 \label{ehDropForm}
\end{figure}
The results of numerical calculations in \figref{ehDropForm} show both the
surface tension and surface thickness of the drop tend to constant values as the
drop size increases. For large drops the boundary becomes approximately flat so
the surface thickness becomes independent of drop radius, as does the surface
tension since its major contribution comes from the drop boundary. We now
examine the surface thickness and tension more carefully in turn.

To derive an approximate expression for the surface thickness we use the
analytical approximation \eqnref{eqn:ehDropDensityInner} to the density profile
of the inside of the drop, which gives the density reduction from the drop
center
\begin{equation}
 \Delta n(r)=-\frac{4\mu\bar{n}}{Q^{3/2}r}\e{Qr}.
\end{equation}
From this, the surface thickness $D$ over which density falls from $90\perc$ to
$10\perc$ is given in the large drop limit $r_{\text{m}}\gg D$ by
\begin{equation}
 D\approx\frac{\ln9}{Q}.
 \label{eqn:DropSurfaceThickness}
\end{equation}

In the given example in \figref{ehDensityProfile} (12 flavors) this predicts
that the surface thickness is $D\approx0.8a_{0}^{*}$, which is of similar size
to the result found by numerical solution of \eqnref{eqn:dropdiffeqn} of
$\sim1.1a_{0}^{*}$, but indicates that the approximation for surface thickness
in \eqnref{eqn:DropSurfaceThickness} is not able to produce accurate results.
The values for surface thickness of drops found using the many-flavor theory can
be compared with results from other approximations. For the six flavor gas in
the large drop limit the many-flavor theory approximation
\eqnref{eqn:DropSurfaceThickness} predicts a thickness of $1.2a_{0}^{*}$, and
exact numerical integration of the many-flavor theory \eqnref{eqn:dropdiffeqn}
predicts thickness $1.6a_{0}^{*}$. The silicon six
flavor result of \citet{78kv03} has a surface thickness of
$1.6a_{0}^{*}$, which is in good agreement with the many-flavor result.

\begin{figure}
 \includegraphics[width=\columnwidth]{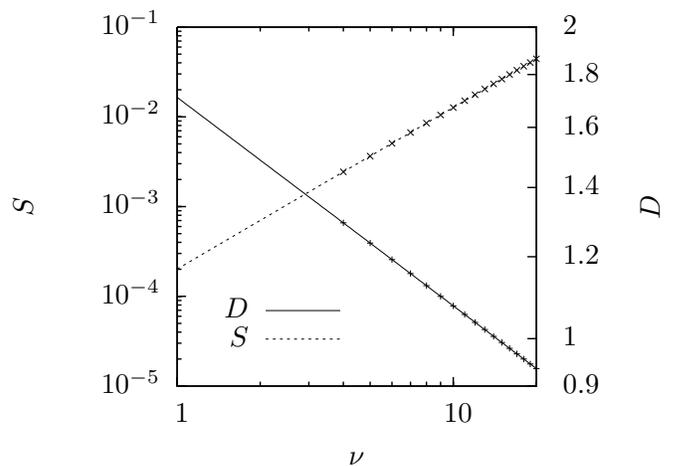}
 \caption{The variation of surface tension (dashed line and crosses) and surface
 thickness (solid line and pluses) with number of flavors present, the straight
 line fits are used to give the exponents of the scaling parameters.}
 \label{ehScaling}
\end{figure}

Having used many-flavor theory to predict the density profile and surface
thickness of an electron-hole drop, it is interesting to examine their scaling
relationships with number of flavors. This is because the scaling relationships
can be experimentally probed \cite{74ps07} by comparing the surface thickness
before and after putting the material under a strain which reduces the valley
degeneracy, for example in silicon from six to two flavors. These scaling
relations will also allow the many-flavor results to be further compared with
previous theoretical work. In \secref{sec:FreeCoulombGas} it was shown that the
expected MFEG uniform density is $\bar{n}\propto\nu^{8/5}$, and from
\eqnref{eqn:ehDropQ2Defn} $Q^{2}\propto\nu^{2/5}$, which with
\eqnref{eqn:DropSurfaceThickness} predicts surface thickness to scale as
$D\propto\nu^{-1/5}$. This scaling prediction for surface thickness can be
compared with numerical results for the variation of surface thickness with
number of flavors in \figref{ehScaling}, found by solving the differential
equation \eqnref{eqn:dropdiffeqn}.  The coefficient for surface thickness
$D\propto\nu^{\alpha}$ is $\alpha=-0.19995(7)$ in good agreement with the
predicted value of $-1/5$.  We can also qualitatively compare our scaling result
with numerical results \cite{78kv03,79rly08} from studies of the electron-hole
drop in silicon. These studies compared results for silicon found at the
unstrained six flavor with the results at two flavors to attempt to model the
effect of stress reducing valley degeneracy. Though two flavor calculations
cannot be accurately given by the many-flavor theory, the qualitative variation
of surface tension and surface thickness should be. The variation of surface
thickness with number of flavors $D\propto\nu^{-1/5}$ is weak, for silicon from
six to two flavors the many-flavor theory, assuming it is valid, predicts that
the thickness increases by a factor of $1.2$. This compares reasonably with the
numerical results of Ref.~\cite{78kv03}, which predicts an increase in surface
thickness by a factor of $\sim1.1$.

The dominating contribution to surface energy is at the boundary of the drop so
the surface tension in large drops is approximately the gradient term in the
energy density, \eqnref{eqn:ManyFlavorGradientExpansion} (the main contribution
to the surface tension) multiplied by the surface thickness $D$
\begin{equation}
 \gamma\approx\frac{(\nabla n)^{2}}{8\bar{n}}D\approx\frac{\bar{n}}{8D},
\end{equation}
where we use the additional approximation $\nabla n\approx\bar{n}/D$. Finally,
with the relationship found above, $D\propto\nu^{-1/5}$, and
$\bar{n}\propto\nu^{8/5}$ found in \secref{sec:FreeCoulombGas}, this predicts
surface tension varies with number of flavors as $\gamma\propto\nu^{9/5}$. The
numerical results of \figref{ehScaling}, found by solving the differential
equation \eqnref{eqn:dropdiffeqn} exactly, predict a coefficient for
$\gamma\propto\nu^{\alpha}$ of $\alpha=1.8004(3)$ in good agreement with the
analytical result, $9/5$. For silicon, reducing the number of flavors from six
to two, the above result predicts that surface tension reduces by a factor of
$7$. This qualitatively agrees with the variation seen by
Refs.~\cite{78kv03,79rly08} of a reduction by a factor of $3$, though comparison
is difficult due to the presence of holes and there being too few flavors
present for the many-flavor theory to be fully applicable.

\section{Conclusions}

This paper describes a new formalism for calculating the behavior of a MFEG. In
the many-flavor limit the Fermi momentum reduces as
$p_{\text{F}}\propto\nu^{-1/3}$ so is small compared with the momenta associated
with the strongest interactions. Intra-valley interactions are more significant
than inter-valley.

The behavior of a homogeneous MFEG in the limit of many-flavors was
derived. Specifically the exact interacting energy per particle is
$E_{\text{int}}=-0.574447(E_{\text{h}}^{*}a_{0}^{*3/4}m^{*1/4})n^{1/4}$; making
it energetically favorable for the MFEG to be dense. The formalism was found to
apply with as few a six flavors over the density range $0.03\nu\ll
na_{0}^{*3}\ll0.005\nu^{4}$.

The MFEG has short characteristic length-scales which motivates a LDA. A
gradient expansion of the energy density with the lowest order term $|\nabla
n|^{2}/8n$ was derived, which was applied to electron-hole drops to study their
density profile and surface properties. Surface thickness was found to scale as
$D\propto\nu^{-1/5}$, surface tension as $\gamma\propto\nu^{9/5}$.

It would be useful to compare our analytical results with those from computer
simulations to verify our findings for the uniform MFEG, its polarizability and
the gradient expansion. This would allow the limits over which the many-flavor
limit applies to be derived more accurately, and allow the formalism to be
applied to more physical systems.

\acknowledgements GJC is grateful to Peter Haynes for useful discussions, Andrew
Morris for careful reading of the manuscript, and acknowledges the financial
support of the EPSRC.


\end{document}